\documentstyle[10pt]{article}

\def\bfl{\begin{flushleft}}
\def\efl{\end{flushleft}}
\def\bfr{\begin{flushright}}
\def\efr{\end{flushright}}
\def\bc{\begin{center}}
\def\ec{\end{center}}
\def\be{\begin{equation}}
\def\ee{\end{equation}}
\def\ba{\begin{eqnarray}}
\def\ea{\end{eqnarray}}
\def\nn{\nonumber }

\def\text#1{\mbox{#1}}
\def\drm{\text{d}}
\def\mevfcub{\;\text{MeV} \, \text{fm}^{-3}} 
\def\ergccub{\;\text{erg} \, \text{cm}^{-3}}
\def\Sign#1{\, \text{sign}\left[#1\right] }
\def\Adequa{\Longleftrightarrow}
\def\PtoMterm{ \left\{ ``-" \right\} }
\def\Vacen#1{E_{v #1}}
\def\Vaceneq#1{E_{v #1}^0}

\begin{document}
~~\\
~~\\
~~\\
~~\\
~~\\
~~\\
~~\\
\bfr
Int. J. Mod. Phys. D 8 (1999) 363-371 \\
gr-qc/9802021\\
\efr

\bc
{\LARGE \bf
Singular shells of quark-gluon matter
}

~~\\
{\large Konstantin G. Zloshchastiev}\\
~~\\
Metrostroevskaya 5/453, Dnepropetrovsk 320128, Ukraine.\footnote{Email: 
zlosh@email.com }
\ec

~\\

\abstract{
The spherically symmetric thin shells of the macroscopically stable 
quark-gluon matter 
are considered within the frameworks
of the bag model and theory of discontinuities in general relativity. 
The equation of state for the two-dimensional matter is suggested, and its
features are discussed.
The exact equations of motion of such shells are obtained.
Distinguishing the two cases, circumstellar and microscopical shells, 
we calculate the parameters of equilibrium configurations,
including the conditions of decay (deconfinement).
}

~\\

PACS number(s):  11.27.+d, 12.38.Mh, 97.10.Fy\\

~~\\
~~\\
~~\\

\newpage
\large

In a lot of models of strong interactions the strange quark matter,
consisting of the $u$-, $d$-, $s$-quarks, 
is the most energetically favourable state   
of baryon matter.
Because of small electrical charge of the quark component as well as 
due to the saturation property, analogous to that of nuclear forces, 
the strange quark bags with the baryon number, varying within the wide 
ranges $10^2 - 10^{57}$, are stable.
Witten \cite{wit} specified the two ways of formation of the 
quark-gluon matter (QGM):
the quark-hadron phase transition in the early Universe and conversion of 
neutron stars into the strange ones.
Quark bag models in the theories of strong interactions 
(the MIT \cite{cjjtw}, IETP, chiral models) suppose 
that the breaking of physical vacuum takes place inside hadrons.
As a result, the vacuum energy densities outside and inside a hadron become 
essentially different.
Appearing in this case the vacuum pressure $B$ on a bag wall 
equilibrates the pressure of quarks and thus stabilises the system.
Phenomenological estimations give $B = (50 - 91.5) \mevfcub$.
In accordance with the quantum chromodynamics (QCD) and 
properties of the quark-gluon plasma (QGP) \cite{cp,env} it is 
usually 
supposed that the bag surface consists of such a quark-gluon plasma, that is 
the lightest $u$-, $d$-quarks and gluons, which were the first to come to the 
deconfinement state.
Thermodynamically the bag is described by the limit equation 
of state \cite{cjjtw,sz}
\be
^{(3)}\!p =\frac{1}{3} \left(\,  ^{(3)}\!\varepsilon - 4 B \right),
\label{eq1}
\ee
where $^{(3)}\!\varepsilon$ and $^{(3)}\!p$ are the total energy density
 and pressure inside the bag respectively,
$4 B = \varepsilon_{\text{QGP}}$
is the energy density of the quark-gluon plasma (if one supposes 
$B=B_{\text{MIT}} = 67 \mevfcub$ then 
$\varepsilon_{\text{QGP}} = 4.3\times 10^{35} \ergccub$).
The bag constant $B$ determines a value of the confinement strength.
The equation of state (\ref{eq1}), in which quarks and gluons are considered
as free massless particles, demonstrates the agreement between QCD and 
macroscopical properties of stable QGP determining the macroscopical
quark-gluon matter.

Before we start let us remark upon the used system of units.
In our theory we have three constants: 
the velocity of light $c$, the gravitational constant $\gamma$ and the
characteristic length of confinement $l_c \approx 1 \,\text{fm}$.
It can easily be seen that the system of these units has the same 
completeness  
as the Planckian one,  i.e., any value can be made dimensionless.
Therefore, below we will consider all the quantities as dimensionless ones, 
except the cases when numerical estimates will be represented.

Thus, the aim of present paper is to study the thin spherically 
symmetric shells of the macroscopically stable quark-gluon matter  
within the frameworks of 
general relativity, thereby it will be shown below
that inclusion of such shells makes the bag model assumptions viable from
the viewpoint of general relativity.
The singular thin shell turns to be a model of some independent shell-like 
entities
(for instance, the shell rejected by the strange star or the surface of a
phase transition), whose thickness is negligible in comparison with a 
circumference radius.
It is described by a three-dimensional closed hypersurface, embedded in
the four-dimensional spacetime and dividing it into the two domains, 
the external ($\Sigma^+$) and internal ($\Sigma^-$) spacetimes.
Beginning from \cite{dau,isr} the theory of surface layers has been
widely considered in the literature 
(see ref.\ \cite{mtw} for details)\footnote{Nevertheless, 
the features of singular layers as models of particle-like objects
remain to be relatively unknown, see ref.\ \cite{zlop} and 
references therein.}.
We only point out the vital difference between the description of the 
boundary surface (for instance, the surface of star) and shell.
The first one is the discontinuity of the first kind 
(the density has a finite jump across the surface) 
and is described by the Lichnerowicz-Darmois junction conditions
(the first and second quadratic forms are continuous on the surface).
The thin shell is the discontinuity of the second kind
(the density has the delta-like singularity on the shell) and is 
described by the Lichnerowicz-Darmois-Israel junction conditions:
the first quadratic form is continuous, the second one
(the extrinsic curvature) has a finite jump.

So, one considers a thin shell with the surface stress-energy 
tensor of a perfect fluid in the general case 
\be
S_{ab}=\sigma u_a u_b + p (u_a u_b +~ ^{(3)}\!g_{ab}),
                                                               \label{eq3}
\ee
where $\sigma$ and $p$ are respectively the surface energy density and 
pressure, $u^a$ is the timelike unit tangent vector, 
$^{(3)}\!g_{ab}$ is the 3-metric on a shell.
We suppose the metrics of the spacetimes outside $\Sigma^+$ and inside 
$\Sigma^-$ of a spherically symmetric shell to be in the form
\be
\drm s_\pm^2 =
-[1+\Phi^{\pm}(r)] \drm t^2_\pm + [1+\Phi^{\pm}(r)]^{-1} \drm 
r^2 + r^2 \drm \Omega^2,                                       \label{eq4}
\ee
where $d\Omega^2$ is the metric of the unit 2-sphere.
It is possible to show that if one uses the  proper time  
$\tau$, then the 3-metric of a shell is
\be
^{(3)}\!\drm s^2 = - \drm \tau^2 + R^2 \drm \Omega^2,        \label{eq5}
\ee
where $R=R(\tau)$ is a proper radius of a shell.
Also the energy conservation law for the matter on a shell can be
written as
\be
\drm \left( \sigma ~^{(3)}\!g \right) +
p~ \drm \left( ~^{(3)}\!g \right) 
+ ~^{(3)}\!g~  \Delta T^{\tau n}\, \drm \tau =0,               \label{eq6}
\ee
where $\Delta T^{\tau n} = (T^{\tau n})^+ - (T^{\tau n})^-$,  
$T^{\tau n}=T^{\alpha\beta} u_\alpha n_\beta$ is the
projection of the stress-energy tensors in the $\Sigma^\pm$
spacetimes on the tangent and normal vectors, $^{(3)}\!g=\sqrt{-
\det{(^{(3)}\!g_{ab})}} = R^2 \sin{\theta}$.
In this equation, the first term corresponds to a change in
the shell's
internal energy, the second term corresponds to the work done
by the shell's internal forces, while the third term
corresponds to the flux of the energy through the shell.

Imposing junction conditions across the shell,
we derive the equations of motion of such shells in the form
\be
\epsilon_+ \sqrt{1+\dot R^2+\Phi^+(R)} - \epsilon_-
\sqrt{1+\dot R^2+\Phi^-(R)} = - \frac{m(R)}{R},            \label{eq7}
\ee
\be
m(R) \equiv 4 \pi \sigma R^2,                            \label{eq8}
\ee
where $\dot R=\drm R/\drm\tau$ is a proper velocity of the
shell, $\epsilon_\pm = \Sign{\sqrt{1+\dot R^2+\Phi^\pm
(R)}}$, $m (R)$ is interpreted as the (effective) rest mass.

Equations (\ref{eq6}) - (\ref{eq8}) together 
with the equation of state $p=p(\sigma,R)$, choice of the 
metrics $\Phi^\pm(r)$ (\ref{eq4}) and signs $\epsilon_\pm$, 
completely determine the motion of the shell in 
general relativity.
Therefore, we must resolve the next three problems:
(i) the choice of $\epsilon_\pm$ in eq.\ (\ref{eq7}), 
(ii) the choice of an equation of state, 
(iii) the choice of $\Phi^\pm(r)$.

(i) {\it The choice of signs $\epsilon_\pm$}.
It is well-known, that $\epsilon = +1$ if $R$ increases in 
the outward normal of the shell ( e.g., it 
takes place in a flat spacetime),
and $\epsilon = -1$ if $R$ decreases (a semiclosed world).
Thus, only under the additional condition $\epsilon_+ = 
\epsilon_-=1$ we have an ordinary shell \cite{bkt}.
In the paper we will deal purely with such shells.

(ii) {\it The choice of the equation of state} of the two-dimensional matter.
For the thick layer (bag) we already have the phenomenological equation
of state (\ref{eq1}).
Now we attempt to connect it with the (spatially) two-dimensional equation 
of state, which would take into account all features of eq.\ (\ref{eq1}).
The correspondence between the integrals
\be
\int \,^{(3)}\!\varepsilon~ \drm
\left( 4 \pi R^3 / 3 \right)
\Adequa
\int \sigma ~\drm\left( 4 \pi R^2\right)                   \label{eq9}
\ee
is a correct way for it.
Hence we have
\be
\sigma \Adequa 2~^{(3)}\!\varepsilon R,~~
p \Adequa 2 ~ ^{(3)}\!p R,                                 \label{eq1011}
\ee
and, therefore, we can connect eq.\ (\ref{eq1}) with the equation of state
\be
2 p = \sigma - 8 B R,                                      \label{eq12}
\ee
taking into account the reduction of dimensionality 
of the stress-energy tensor.
It should be noted that this equation has not to be the equation of 
state for the bag surface if it assumed to be the boundary surface.
The boundary surfaces (i.e., discontinuities of first kind) can not have
{\it a priori} neither surface density of energy nor surface pressure.
As for the QGM layer that it has to be the independent object in 
which quarks are trapped.
In the thin-wall approximation it is described by a singular hypersurface,
and the relation (\ref{eq12}) turns to be the effective
equation of state of the two-dimensional strange matter, quark-gluon matter.
Thus, the physical picture is as follows.
Inside the quark bag the QCD vacuum is degenerated, quarks are highly 
compressed into a radiation fluid and turn to be free particles;
the principle of asymptotical freedom says that strong forces increase 
when distance between quarks increases. 
Further, on the surface of a bag 
the pressure decreases, hence
the distance between quarks increases that leads to an appearance of the
strong gluon interaction and stable quark-gluon composite.
 
Let us find both the surface density and pressure as explicit functions 
of radius.
They are determined by the solution of the differential equation (\ref{eq6}) 
with provision for both eq.\ (\ref{eq12}) and 
$(T^{\tau n})^+ = (T^{\tau n})^- = 0$.
The last condition is valid for all the metrics (\ref{eq4})
(see \cite{zlo} and references therein).
We have
\be
\sigma / 2 =  B R + \delta^2 R^{-3},~~
p = - 3 B R + \delta^2 R^{-3}.                      \label{eq1314}
\ee
It is evident that the QGM shell is not the bubble, for which
$\sigma = -p = \text{constant}$.
These expressions correspond, according to eqs.\ (\ref{eq1011}), to
\be
^{(3)}\!\varepsilon =  B  + \delta^2 R^{-4},~~
^{(3)}\!p = - B + \delta^2 R^{-4}/3.                \label{eq1516}
\ee
Here $\delta$ is the integration constant, whose value and sense will 
be found below.
We call it the {\it quasicharge of the matter}, since, it
(a) determines a value of the effective repulsion force between QGM 
particles, 
(b) has a dimensionality of charge.
Supposing $^{(3)}\!p = 0$ at $R=1$, we obtain
\be
\delta^2 = 3 B,                                             \label{eq17}
\ee
and hence we have the expected result 
$ ^{(3)}\!\varepsilon |_{R=1}= 4 B = \varepsilon_{\text{QGP}}$.          
It can readily be seen that at $R>1$ the pressure is negative, and 
repulsion forces are relayed by the attractive ones, thus counteracting  
to infinite expansion (deconfinement, see below) of the shell.
At $R<1$ the pressure, on the contrary, prevents from unlimited collapse.

As an addition we give the proof by contradiction that the true QGM shell  
must always have a finite radius.
Let us suppose, for instance, that both eqs.\ (\ref{eq1516}) 
and equation of state (\ref{eq1}) are satisfied.
Then at $R = \infty$
\be
^{(3)}\!\varepsilon_\infty = B,\,
^{(3)}\!p_\infty = -B,                                        \label{eq18}
\ee
and, therefore, the asymptotic equation of state 
\be
^{(3)}\!\varepsilon_\infty + ^{(3)}\!p_\infty = 0           
\label{eq19}
\ee
does not coincide with eq.\ (\ref{eq1}), Q.E.D.
This is the key difference of the QGM shell from the cases studied earlier, 
namely, the dust shells, bubbles, several fluid shells.
Those  can formally have an infinite radius, because they preserve a proper 
equation of state and thus their own existence at infinity.

(iii) {\it The choice of metrics} of spacetimes outside 
and inside the QGM shells.
The choice of fixed $\Phi^\pm (r)$ must take 
into account at least the two physical factors, namely, 
the electrical neutrality of QGM 
and existence of the nontrivial QCD vacuum.
The estimate of energy density of the physical vacuum   
with respect to the vacuum of the perturbation theory was obtained 
by means of the QCD sums 
in ref. \cite{nsvz}, see also ref. \cite{nar}
\be
\varepsilon_v = - 500 \mevfcub.                             \label{eq20}
\ee
In general relativity a physical vacuum  is usually considered through 
the application of the de Sitter term \cite{his}.
Thus, taking into account the aforesaid, we suggest
\be
\Phi^\pm (R) = - 2 M_\pm/R - 8 \pi \varepsilon_{v \pm} R^2/3,                                 \label{eq21}
\ee
which corresponds to the neutral QGM shell surrounding a neutral QCD object 
( e.g., strange star or neutral hadron) with mass $M_-$.
Then $M_+$ is the total observable mass (energy) of the whole configuration.

Finally, taking into account the items (i)-(iii), the equations of 
motion of the QGM shells (\ref{eq7}) can be written more strictly
\be
\sqrt{
      1-\frac{2 (M_+ + \Vacen{+})}{R}  
+ \dot R^2
       } -
\sqrt{
      1-\frac{2 (M_- + \Vacen{-})}{R}  
+ \dot R^2
     }
= -\frac{m(R)}{R},                                         \label{eq22}
\ee
\be
m(R) = 8 \pi (B R^4 + \delta^2) /R
= 8 \pi B (R^4 + 3) / R,                                \label{eq23}
\ee
where $\Vacen{\pm} = 4 \pi \varepsilon_{v\pm} R^3 /3$.
From eq.\ (\ref{eq22}) it can readily be seen that the sufficient conditions 
of matching  of the spacetimes $\Sigma^+$ and $\Sigma^-$ across the shell 
at any $R$ are the inequalities
\be
M_+ > M_-,~~                                                   
\varepsilon_{v+} > \varepsilon_{v-}.                       \label{eq2425}
\ee
In this connection the following fact is attractive.
If one supposes the external metric (\ref{eq4}) to be asymptotically flat
(the shell completely screens the QCD vacuum from an observer in $\Sigma^+$ 
and it corresponds to the modern observable physical picture), 
then $\varepsilon_{v+} = 0$ and hence $\varepsilon_{v-} < 0$.
Thus, the expected sign of the vacuum energy is in agreement with the 
estimate (\ref{eq20}) obtained irrespective of general relativity.

Further we study the static QGM shells.
Differentiating eq.\ (\ref{eq22}) with respect to $\tau$, we obtain
\be
\frac{
      \ddot R + (M_+ - 2 \Vacen{+})/R^2 
     }
     {
      \sqrt{
            1- \frac{2 (M_+ + \Vacen{+})}{R} 
            + \dot R^2
           }      
     }
- \PtoMterm = - \left(\frac{m(R)}{R}\right)_{,R},            \label{eq26}
\ee
where $\PtoMterm$ means the term obtained from the first term by 
replacing the subscript ``$+$'' on $M$ and $E_v$ 
by ``$-$''.
Equations (\ref{eq22}) and (\ref{eq26}) at $\dot R = \ddot R =0$  give 
the equilibrium conditions \cite{fhk}
\ba
&&\sqrt{
         1-\frac{2 (M_+ + \Vaceneq{+})}{R_0}  
       } - \PtoMterm = - \frac{m_0}{R_0}, \nn\\
&&                                                           \label{eq27}\\
&& \frac{ M_+ - 2 \Vaceneq{+} 
        }
        {
          \sqrt{
                1-\frac{2 (M_+ + \Vaceneq{+})}{R_0} 
               } 
        } 
- \PtoMterm = -  R_0^2
\left(\frac{m_0}{R_0}\right)_{,R_0},\nn
\ea
where $\Vaceneq{\pm} = \Vacen{\pm}|_{R=R_0}$, $m_0 = m (R_0)$.
Unfortunately, exact analytical examination of the system
(\ref{eq27}) is impossible because of mathematical troubles.
So we perform expansion in series with respect to 
the small parameter $\varepsilon_{v-}$ (according to eq. (\ref{eq20}),
dimensionless $|\varepsilon_{v-}| \sim 10^{-19}$).
Then the system (\ref{eq27}) yields
\ba
&& \frac{
         \alpha_+^2 - \Vaceneq{+}
        } 
        {\alpha_+} 
- \PtoMterm = - \frac{m_0}{\sqrt{R_0}}, \nn\\
&&                                                          \label{eq28}\\
&&\frac{ \alpha_+^2
         \left( 
               M_+  - 2 \Vaceneq{+}
         \right) +
         M_+ \Vaceneq{+}
        }
        {\alpha_+^3 R_0^{3/2}
        }
-  \PtoMterm
= -  \left(\frac{m_0}{R_0}\right)_{,R_0},\nn
\ea
where $\alpha_\pm = \sqrt{R_0 - 2 M_\pm}$.

Further we consider the two approximations corresponding to the
microscopic (a) and astronomical (b) QGM shells.

(a) {\it M-approximation}.
It is well-known that radii of elementary particles are
much greater than their horizon radii, hence $R_0 \gg 2 M_\pm$.
Then to the first order in 
$2 M_\pm/R_0$ the system (\ref{eq28}) can be greatly simplified.
Taking into account eq.\ (\ref{eq23}), we have
\be
\Delta \varepsilon_v = 6 B (1-R_0^{-4}),~ \Delta M = 32 \pi B 
R_0^{-4}, \label{eq30}
\ee
where 
$\Delta \varepsilon_v =\varepsilon_{v+} - \varepsilon_{v-}$,
$\Delta M =M_{+} - M_{-}$.
Solving this system, we obtain both the radius of the microscopic QGM 
shell in equilibrium
\be
R_0 = B^{1/4}\,\left( B-B_c\right)^{-1/4}                \label{eq31}
\ee
and the value of the mass amplification effect \cite{zlo}
\be
\Delta M = 32 \pi B^{3/4} (B-B_c)^{1/4}, \label{eq32}
\ee
where the critical value of the bag constant is
\be
B_c = \Delta \varepsilon_v/6.                             \label{eq33}
\ee
Hence it can readily be seen that at $B\leq B_c$ the decay of QGM shells 
takes place (see in this connection remarks (\ref{eq18}), (\ref{eq19})).

The quantitative estimations will be performed upon assuming of
the full screening of QCD vacuum,  i.e.,
$\varepsilon_{v+}=0$ (see the comments after eq.\ (\ref{eq2425})).
Then taking into account eq.\ (\ref{eq20}), we obtain
\be
B_c \approx 83.33 \mevfcub.                                  \label{eq34}
\ee
Finally, following eqs.\ (\ref{eq31}), (\ref{eq32}), and 
(\ref{eq34}), we represent the values of $R_0$ and $\Delta M$ for some 
chosen bag constants $B$ (in $\mevfcub$)
\ba
B=83.334: && R_0=18.8,  \, \Delta M=0.48 m_p, \nn\\
B=86.7:   && R_0=2.25,  \, \Delta M=4.13 m_p, \nn\\
B=91.5:   && R_0=1.83,  \, \Delta M=5.37 m_p, \nn
\ea
where $m_p$ is the proton rest mass.
Further, $\Delta M$ does not depend on $M_\pm$ (\ref{eq32}) and, 
therefore, can be interpreted as the proper mass
of an equilibrium QGM shell with provision for gravity.
In this connection it should be pointed out that the estimations above
do not speak (almost) nothing about internal structure of such shells.
For instance, $\Delta M = m_{n^0}$ (at some $B$)
does not mean that the shell consists of a single neutron 
and thus models it.
These shells are initially defined to be macroscopical 
entities (and evidently
cannot be used for modelling the known particles)
but their masses can be
less than the sum of masses of constituent particles because of
both the strong binding energy and gravitational defect of mass,
thereby the latter is non-negligible as was shown.
Besides, one can see that in this case 
admissible $B$ should be more than 
$83.33 \mevfcub$ that, as it is well-known, corresponds to 
macroscopical bags
rather than to those modelling the known non-resonance particles 
for which $B \sim 60-70 \mevfcub$.

Finally, of course these estimates can be 
sufficiently conditional, since, in the real microscopical world the model 
of the (globally) classical spherical QGM shell can be too 
rough because of presence of uncompensated spins, 
electrical charges, quantum 
fluctuations, pair creations etc.

(b) {\it A-approximation}.
Now we consider the case of the shells with great masses and radii.
It means that 
(i) we neglect the value of mass amplification 
($\Delta M/M_\pm \ll 1$), 
(ii) the shell radius $R \gg 1$, hence the repulsion 
(i.e., the quasicharge $\delta$) in eq.\ (\ref{eq23}) can be neglected.
Thus, we assume $M_+=M_-=M$, $\delta=0$.
In addition, we assume 
the full screening of the QCD vacuum $(\varepsilon_{v+}=0)$.
Then eqs.\ (\ref{eq22}) and (\ref{eq26}) after expansion in series with 
respect to $\varepsilon_{v-}$ yield, respectively,
\ba
&&\sqrt{1- 2 M/R +\dot R^2} = B_c/B,                \label{eq35}\\
&&\dot R \left( \ddot R + M/R^2 \right) = 0,        \label{eq36}
\ea
where $B_c$ is determined by eq.\ (\ref{eq33}) 
at $\Delta \varepsilon_v \equiv -\varepsilon_{v-}$ 
and evaluated in eq.\ (\ref{eq34}).
One can see that the last equation is
satisfied identically at equilibrium, and eq.\ (\ref{eq35}) yields
\be
R_0 = 2 M  B^2 \left( B^2-B_c^2 \right)^{-1}.                 \label{eq37}
\ee
Notice, this expression has the physical sense only at $B > B_c$, 
then always $R_0 > 2 M$.
In other words, the critical value $B_c$ plays its role at astronomical 
scales too.
Finally, we make some quantitative estimations.
Let the QGM shell be surrounding the strange star considered
by Witten ($M_{\text{star}}=1.63 M_{\odot}$, $B=86.7
\mevfcub$, $R_{\text{star}}=8.82\,\text{km}$).
Strictly speaking, we could assume any other, suitably small, 
massive star (e.g., the neutron star) instead of the strange 
one, but the strange star seems to be the most probable astronomical 
source of quark-gluon matter.
Besides, we will assume that the star and enveloping shell 
are ``made'' from the same matter, i.e. their bag constants are equal.
Then, according to eq.\ (\ref{eq37}), the radius of an
equilibrium QGM shell has to 
be $R_0 \approx 32\,\text{km}$, i.e., more than $3.6$ radii of the star.

Thus, in the present paper some aspects of dynamics of the spherically
symmetric quark-gluon matter shell  were considered.
Let us summarise briefly the main results obtained.
The general physical sense of the QGM shells is the shell appears to be
a natural mechanism of the cut-off of vacuum energy.
Indeed, the de Sitter term (widely used for describing of vacuum energy)
is non-vanishing (moreover, increases as $R^2$) at infinity, 
hence we have the disagreement with 
the known properties of strong interactions at large distances.
Thus, the QGM shell obviates the contradiction between the global nature
of the de Sitter term and the short-range nature of the strong forces.
Further, we have obtained not only the exact relativistic equations of 
motion and conditions of equilibrium of the QGM shells 
but also the conditions of their decay in terms 
of the vacuum energy.
Thereby, the critical value of the bag constant $B_c$ appears both as the
additive restriction for $B$ and the useful parameter for the
detection of the possible nucleation-decay of QGM shells.

Of course, we supposed these shells to be globally classical entities.
On the other hand, it would be very important to study 
their quantum properties, including possible part of such shells
at the quark-hadron phase transition in the early Universe \cite{rm}.

\def\CMPh{Commun. Math. Phys.}
\def\JPh{J. Phys.}
\def\CJP{Czech. J. Phys.}
\def\FP{Fortschr. Phys.}
\def\IJMP  {Int. J. Mod. Phys.}
\def\LMPh {Lett. Math. Phys.}
\def\MPL  {Mod. Phys. Lett.}
\def\NPh  {Nucl. Phys.}
\def\PhE  {Phys. Essays}
\def\PhL  {Phys. Lett.}
\def\PhR  {Phys. Rev.}
\def\PhRL {Phys. Rev. Lett.}
\def\PhRp {Phys. Rep.}
\def\NCim {Nuovo Cimento}
\def\NuPB {Nucl. Phys.}
\def\GRG {Gen. Relativ. Gravit.}
\def\CQG {Class. Quantum Grav.}
\def\prp {report}
\def\Prp {Report}

\def\jn#1#2#3#4#5{{\it #1}{#2} {\bf #3}, {#4} {(#5)}}
\def\boo#1#2#3#4#5{{\it #1} ({#2}, {#3}, {#4}){#5}}
\def\prpr#1#2#3#4#5{{``#1,''} {#2}{ #3}{ No. #4}, {#5} (unpublished)}

\end{document}